\input pipi.sty
\input epsf.sty
\input psfig.sty

\raggedbottom
\nopagenumbers
\rightline\timestamp
\rightline{FTUAM 06-19}
\rightline{hep-ph/0701025}

\bigskip
\hrule height .3mm
\vskip.6cm
\centerline{{\bigfib Experimental status of the $\pi\pi$ isoscalar S wave at low energy:}}
\medskip
\centerline{{\bigfib $f_0(600)$ pole and
scattering length}}
\medskip
\centerrule{.7cm}
\vskip1cm

\setbox9=\vbox{\hsize65mm {\noindent\fib and F. J. 
Yndur\'ain} 
\vskip .1cm
\noindent{\addressfont Departamento de F\'{\i}sica Te\'orica, C-XI\hb
 Universidad Aut\'onoma de Madrid,\hb
 Canto Blanco,\hb
E-28049, Madrid, Spain.}\hb}

\setbox8=\vbox{\hsize65mm {\noindent\fib R. Garc\'{\i}a-Mart\'{\i}n and J. R. Pel\'aez} 
\vskip .1cm
\noindent{\addressfont Departamento de F\'{\i}sica Te\'orica,~II\hb
 (M\'etodos Matem\'aticos),\hb
Facultad de Ciencias F\'{\i}sicas,\hb
Universidad Complutense de Madrid,\hb
E-28040, Madrid, Spain}}
\bigskip
\centerline{\box8}
\bigskip
\centerline{\box9}
\bigskip
\setbox0=\vbox{\abstracttype{Abstract} The experimental results obtained in the last few
years on kaon decays (K$\to2\pi$ and, above all, Ke4 decays) 
allow a reliable, model independent determination of low energy $\pi\pi$ scattering in the S0 wave. 
Using them and, eventually, other sets of data, it is possible to give a 
precise parametrization of the S0 wave as well as
to find  the scattering length and effective range parameter.
One can also  perform an  extrapolation
to the pole of the ``$\sigma$ resonance" [$f_0(600)$].
We obtain the results 
$$a_0^{(0)}=0.233\pm0.013\;M^{-1}_\pi,\quad b_0^{(0)}=0.285\pm0.012\;M^{-3}_\pi$$
and, for the $\sigma$ pole,
$$M_\sigma=484\pm17\;\mev,\quad\gammav_\sigma/2= 255\pm10\;{\rm MeV}.$$   
}
\centerline{\box0}

\brochureendcover{Typeset with \physmatex}

\brochureb{\smallsc  
 r. garc\'{\i}a-mart\'{\i}n, j. r. pel\'aez and f. j.  yndur\'ain}
{\smallsc experimental status of
the
$\pi\pi$ isoscalar s wave at low energy: {\petit $f_0(600)$} 
pole and scattering length}{1}
\vskip1truecm

\booksection{1. Introduction.}

\noindent
The lightest scalar and isoscalar mesonic ``resonance", sometimes called the $\sigma$ resonance, 
has enjoyed a peculiar status. The Particle Data
Tables (PDT\ref{1}),  that refer to it as the $f_0(600)$ resonance,
give for its pole a mass between 400 and 1200~\mev, and a half width between 
300 and 500~\mev; and the same values for the 
``Breit--Wigner" mass and half width: wide ranges, indeed. 
  In fact, in the old
days  
 the resonance was many times reported as
non-existent.  On the other hand,  calculations based on chiral perturbation theory
and dispersion relations\ref{2,3}
give the following values for the complex pole corresponding to it:\fnote{The
determinations 
(1.1) and (1.2) are, in fact, based on  different methods. We quote them together 
to give an idea on the spread of {\sl theoretical} calculations that, one way or the other, use 
chiral perturbation theory to get the sigma pole.}
$$
M_\sigma=441\pm17\;\mev,\quad \gammav_\sigma/2=230\pm15\;\mev;\quad\hbox{Ref.~2}
\equn{(1.1)}$$
(the errors here are purely indicative; they are obtained from the spread of  the 
values in the  articles in ref.~2)  and
$$M_\sigma=441^{+16}_{-8}\;\mev,\quad
 \gammav_\sigma/2=279^{+9}_{-12.5}\;\mev;
\quad\hbox{Ref.~3}.
\equn{(1.2)}$$
Similar results are obtained by Zhou et al.,\ref{4} who also use chiral perturbation theory 
and dispersion relations and get
$$M_\sigma=470\pm50\;\mev,\quad  \gammav_\sigma/2=285\pm25\;\mev.
$$
It would be nice to be able to compare these results with {\sl experiment}, without the 
tremendous uncertainties given in the PDT.
Furthermore, it is desirable to use, in the comparison with experiment, only data from the S0
wave at low and intermediate energies 
without introducing  extraneous information on 
the S0 wave at high energies, or large amounts of data from other waves.
This is what is done in the 
present paper, where the only theory we use is unitarity and analyticity, 
the last only what follows
 from general field theory.\fnote{In contrast, the authors of ref.~3 also use debatable theoretical
information (see ref.~5); for example, on the (unmeasurable) scalar form factor of the pion, or on
Regge theory. While in the articles in ref. 2 some theoretical input is used  to control the left hand
cut.}

 In the present note it
will be shown that, indeed, it is possible to get  a determination of $M_\sigma$,
$\gammav_\sigma$, from experimental data on the S0 wave  alone, 
and with a
precision  comparable to the theoretical one in (1.1) or (1.2).   In fact, we will show
that,  {\sl from experiment}, one has the 
values
$$M_\sigma=484\pm17\;\mev,\quad\gammav_\sigma/2= 255\pm10\;{\rm MeV}.
\eqno{(1.3)}$$
The number for $M_\sigma$ is  stable; that for $\gammav_\sigma$ less so.
Thus, theoretical calculations appear to be reasonably compatible with experiment.

Among the low energy 
parameters, the scattering length, $a_0^{(0)}$ and effective range
parameter,  $b_0^{(0)}$, are particularly important. They are defined as in ref.~6,
$$\dfrac{s^{1/2}}{2 M_\pi k^{2l+1}}\real \hat{f}_l^{(I)}(s)\simeqsub_{k\to0}
a_l^{(I)}+b_l^{(I)} k^2+\cdots\,;\quad \hat{f}_l^{(I)}=\sin\delta_l^{(I)}\ee^{\ii\delta_l^{(I)} },
 \quad
k=\sqrt{s/4-M^2_\pi},
$$  
where $\hat f_l^{(I)}$ is the partial wave of definite isospin $I$ and angular momentum $l$, and
$\delta_l^{(I)}$ stands for its corresponding phase shift. 
The same methods that allow us to
extrapolate the  experimental S0 amplitude to the sigma
pole 
permit an extrapolation to find them. The best values we get are
$$a_0^{(0)}=0.233\pm0.013\;M^{-1}_\pi,\quad
b_0^{(0)}=0.285\pm0.012\;M^{-3}_\pi.\equn{(1.4)}$$
These values are very stable; they
 compare well with the results from unitarized chiral perturbation theory 
(cf. the last article in ref.~2)
$$
a_0^{(0)}=0.231^{+0.003}_{-0.006}\;M^{-1}_\pi,\quad
b_0^{(0)}=0.300\pm0.010\;M^{-3}_\pi
\eqno{(1.5)}$$
(with only statistical errors), or with what was found from 
chiral perturbation theory in ref.~7,
$$
a_0^{(0)}=0.220\pm0.005\;M^{-1}_\pi,\quad
b_0^{(0)}=0.276\pm0.006\;M^{-3}_\pi,
\eqno{(1.6)}$$
although the central value of $a_0^{(0)}$ is a bit displaced.

Finally, we can find a parametrization that represents very accurately the S0 wave up to energies
$s^{1/2}\simeq950\,\mev$; see Eq.~(6.1) in the Conclusions below.

The reason why we are able to present such an accurate determination of the
$a_0^{(0)}$,  $b_0^{(0)}$, as well as the location of the
 sigma pole from 
experiment  is the availability, 
in recent years, of very precise data on the S0 wave at low energies, based on K decays, 
and the use of a powerful, model independent fitting technique.

Before embarking on the details of the calculations, however,
 a few words have to be said on the
meaning  of the word ``resonance" in connection with the $\sigma$. 
One can give three definitions of the location of
 an elastic, background free, resonance
mass and width:  the point $s^{1/2}=\mu_0$ at which the phase shift crosses $\pi/2$, and the 
width of the
corresponding  Breit--Wigner parametrization, $\gammav_0$ (usually called 
 ``Breit--Wigner" mass and  width); 
the mass at which the energy derivative of the phase shift is a maximum 
and, for the width, the inverse of such derivative; 
and the pole of the partial wave amplitude in
the  (unphysical) Riemann sheet, at $M_\sigma-\ii \gammav_\sigma/2$.
 Of these, the more physical one
is the second:  it identifies a resonance as a metastable state, 
whose lifetime is the inverse of
the width (Wigner's time delay theory). 

In the case of {\sl narrow} resonances, all three
definitions  coincide to first order in $\gammav/2M$;
 but the situation for the $\sigma$ is very different. 
The  S0 wave phase shift for $\pi\pi$ scattering, $\delta_0^{(0)}(s)$, does indeed cross 
$\pi/2$; but it does so\ref{6} at an energy of $s^{1/2}=\mu_0\simeq800\,\mev$. 
The energy {\sl derivative} of  $\delta_0^{(0)}(s)$ is nowhere maximum near this point,
 and the partial wave amplitude does not resemble a
Breit--Wigner shape. 
Finally, and as we have already advanced in  Eq.~(1.3), there exists a pole 
in the second Riemann sheet, but
at very low energy and with very large width. 
These are the reasons why the  classification of the 
$\sigma$ as a resonance is so controversial. 
At any rate, 
in the present note we will be concerned  with the complex, second Riemann sheet pole 
of the partial wave amplitude, without discussing
 the relevance of the resonance concept for it.

\booksection{2. The resonance condition}

\noindent In order to look for the pole on the second Riemann sheet associated to
the sigma resonance, we will be using the well known fact 
that such a pole, located  at $\sqrt{s_\sigma}=M_\sigma-\ii \gammav_\sigma/2$,
 corresponds exactly with a {\sl zero} of the S-matrix partial wave $S_0^{(0)}(s)$,
$$S_0^{(0)}(s)=1+2\ii \hat{f}_0^{(0)}(s)=\ee^{2\ii\delta_0^{(0)}},$$
 at ${\bar{s}_\sigma}^{1/2}
=M_\sigma+\ii
\gammav_\sigma/2$ in the {\sl physical} Riemann sheet.
 Thus, we can find the location of the
resonance by looking for the solutions of 
$S_0^{(0)}(\bar{s}_\sigma)=0$ in the upper half of the complex plane.
This zero condition may be written in a simpler manner as
$$\cot\delta_0^{(0)}(\bar{s}_\sigma)=-\ii.
\equn{(2.1)}$$ 

 Solving (2.1)  requires analytical continuation from the real axis.
The problem with determining the location of the resonance pole is that
 analytic continuation suffers from 
instability problems, due to the fact that the function
 is not known exactly on the real axis. In the old days, the low
energy experimental data  for the S0 wave were hopeless: several solutions existed, and the
errors  were very large. This is one of  the reasons for the widely different 
determinations of the $\sigma$ pole reported in the PDT.\ref{1}
Fortunately, and as already remarked,  the situation has improved enormously
in the recent years. 
We now have reliable data from 
just above threshold up to the kaon mass, $m_K$, due to 
K$_{l4}$ decays and K$_{2\pi}$ decays.\ref{8} Moreover, and although 
still preliminary, we have even better very recent K$_{l4}$ decay data.\ref{9}

In addition to this, we can profit from the fact that the determinations of 
the S0 wave, although not very precise, are reasonably compatible among themselves 
in the range 810 -- 960~\mev; see ref.~6, Eqs.~(2.13). 
This is sufficient to make a fairly stable analytical extrapolation.

\booksection{3. The conformal mapping method: theory}

\noindent 
The unitarity of the S matrix imposes, in the elastic region, a constraint on 
partial wave amplitudes; for the S0 wave,
$$\imag \hat{f}_0^{(0)}(s)=|\hat{f}_0^{(0)}(s)|^2
$$
If we write 
$$\hat{f}_0^{(0)}(s)=\sin\delta_0^{(0)}(s)
\ee^{\ii\delta_0^{(0)}(s)}=\dfrac{1}{s^{1/2}\psi(s)/2k-\ii}$$
with $k=\sqrt{s/4-M^2_\pi}$ the c.m. momentum, then 
what we know of the analyticity and unitarity properties of the S0 partial wave
amplitude  implies that the function ({\sl effective range 
function})
$$\psi(s)\equiv({2k}/{s^{1/2}})\cot\delta_0^{(0)}(s)$$
 is analytic in the complex plane
 cut from $-\infty$ to 0 and
from the point where inelasticity is important, $s_0$, to $+\infty$ (\fig~1): 
note that the {\sl elastic cut} is absent in $\psi(s)$. 
In the case of the S0 wave, one can take $s_0=4m^2_K$.
\topinsert{
\setbox0=\vbox{{\epsfxsize 10.2truecm\epsfbox{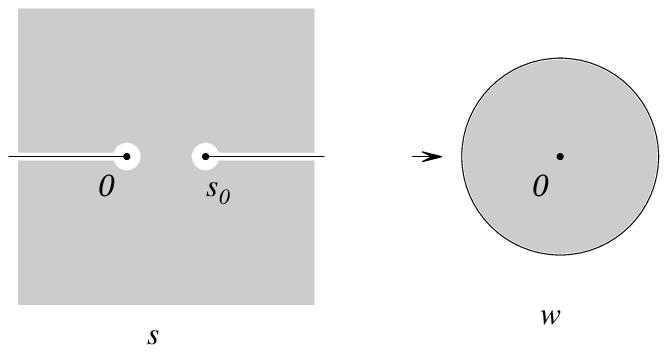}}} 
\setbox6=\vbox{\hsize 6truecm\captiontype\figurasc{Figure 1. }{ The  
mapping $s\to w$.\hb
\phantom{XX}}\hb
\vskip.1cm} 
\medskip
\centerline{\box0}
\centerline{\box6}
\medskip
}\endinsert

The standard mathematical method to deal with this situation is to make 
a conformal mapping, from the variable $s$ to the variable 
$$w(s)=\dfrac{\sqrt{s}-\sqrt{4m^2_K-s}}{\sqrt{s}+\sqrt{4m^2_K-s}}$$
under which the cut plane is mapped into the unit disk (\fig~1) 
and the analyticity properties of the effective range function 
in the complex plane in the variable $s$ are {\sl strictly equivalent} to uniform,
absolute  convergence of a Taylor series for the  function $\psi$ in the variable $w$ 
in the disk $|w|\leq1$. 
For this convergence, it necessary to separate off those 
 poles of the effective range function  that lie on the 
real axis; and it can also be convenient to  separate  the 
zeros  of the effective range function in the same region 
(although this last is not necessary; see below). 
Of these we have one of each: a pole due to the so-called Adler zero of the 
partial wave amplitude, lying near the left hand cut, at
$s=\tfrac{1}{2}z_0^2,$ 
 $z_0\simeq M_\pi$ 
(with $M_\pi$ the charged pion mass); and then there is a zero due to the phase shift 
crossing $\pi/2$ for an energy $s^{1/2}=\mu_0\sim 800~\mev$.
 Thus, we can, in 
all generality, write the following
parametrization [we will at times simplify the notation by writing 
$\delta$ for $\delta_0^{(0)}$]: 
$$\eqalign{
\cot\delta(s)=&{{s^{1/2}}\over{2k}}\,{{M^2_\pi}\over{s-\tfrac{1}{2}z_0^2}}\,
{{\mu^2_0-s}\over{\mu^2_0}}\big\{\hat{B}_0+\hat{B}_1w(s)+\hat{B}_2w(s)^2+\cdots\big\}.\cr
}
$$
However, if using this expression with a finite number of terms, 
it presents the problem that a ghost is generated in the 
partial wave amplitude.\fnote{We are grateful to 
 Profs. Caprini and Leutwyler for this remark.}
 Although this ghost is harmless, being weak and very near 
the left hand cut (see the discussion in Appendix~A) it is, as a matter of principle, better to
use a ghost-free  expansion. For this, it is sufficient to replace the formula above by
$$\eqalign{
\cot\delta(s)=&{{s^{1/2}}\over{2k}}\,{{M^2_\pi}\over{s-\tfrac{1}{2}z_0^2}}\,
{{\mu^2_0-s}\over{\mu^2_0}}\Big\{\dfrac{z_0^2}{M_\pi\sqrt{s}}+
{B}_0+{B}_1w(s)+{B}_2w(s)^2+\cdots\Big\}.\cr
}
\equn{(3.1)}$$

Another possibility is {\sl not} to separate the zero at $s=\mu^2_0$, writing 
simply
$$\eqalign{
\cot\delta(s)=&{{s^{1/2}}\over{2k}}\,{{M^2_\pi}\over{s-\tfrac{1}{2}z_0^2}}\,
\Big\{\dfrac{z_0^2}{M_\pi\sqrt{s}}+{B}_0+{B}_1w(s)+{B}_2w(s)^2+\cdots\Big\}.\cr
}
\equn{(3.2)}$$ 
In these last  case the zero is generated by the combination of the 
$B_nw^n$ and we will, generally speaking,
 need one more term in the expansion than if we used 
(3.1), although the number of free parameters will be the same. 

Now, 
the key point for us is that the expansions
(3.1), (3.2) converge in  the {\sl whole} cut plane: therefore, 
they  can be used 
as they stand to solve (2.1), and hence to find the location of the resonance.
In particular, the expansions converge at threshold and thus can 
also be used to determine the values of the scattering length and effective 
range parameters.  
As to the evaluation of the errors, 
we simply vary the parameters within the errors 
obtained in the fits to data, assuming that they are uncorrelated, 
and thus calculate the errors of the related observables.
We have verified that, in the case of our best fit [Eq.~(4.8) below] 
the parameters are actually almost uncorrelated, 
so this is a valid procedure.

It is also to be noted that, because of the nature of the singularities 
on the cuts, we expect the $B_n$ to decrease like
$$|B_n|\sim (n+1)^{-3/2}
\equn{(3.3)}$$
for large $n$. 
This last condition is, strictly speaking, only valid when one separates off the 
ghost piece, i.e., for the ${B}_n$; but, because the ghost-producing piece 
${z_0^2}/{M_\pi\sqrt{s}}$ is so small, it also would
 hold for the  $\hat{B}_n$ as well.
It should also be taken into account that the 
behaviour (3.3) is expected to set in earlier if we separate out the 
zero, as in (3.1); 
if we do not separate out the zero, the first $B_n$ are
 constrained to cooperate to build up the
zero, so 
the asymptotic regime will set later. 
This is in fact what we observe in our fits.

We can get good fits to data with one or two $B_n$s in (3.1), 
or two to three in (3.2); no more are necessary. 
Generally speaking, the situation is as follows: 
if we include in the fit only {\sl low} energy (K-decay) data, 
then a \chidof\ smaller than unity is achieved with only two parameters. 
If we include a third parameter in the fit, 
there is no substantial decrease of the \chidof, 
and there appear 
spurious minima: hence this third parameter is rejected by the fit.
On the other hand, if we also include in the fit {\sl high} energy 
data, then one can still fit with only two parameters; but if including a third 
parameter, the fit remains meaningful (the \chidof\ decreases appreciably)
and, moreover, it is theoretically more satisfactory. 
In particular, in the case where we fit with expression (3.2),
the presence of terms in $w$, $w^2$ means that we uncorrelate the right 
hand cut and the left hand cut of the partial wave, while with only 
two $B_n$s we would have an {\sl average} representation of both cuts.
But even if we include high energy data, one cannot meaningfully 
fit with four parameters, as spurious minima appear.
All this should be clear from the values of the $B_n$'s and the \chidof's of the various fits in the text.

Another technical point is that we will fix the Adler zero to $z_0=M_\pi$ in our analysis
 (the fit depends very little on the precise value of $z_0$, provided
it is near this), and we let the $B_n$, $\mu_0$ vary.  

Before turning to the actual calculations a few words have to be said on 
the matter of analytical extrapolation. 
In principle, if one knows an analytic function on a segment of the real line, the function is
determined on the whole (cut) plane. 
In practice, however, the function is not known exactly. 
One has, therefore, to test for potential {\sl instabilities} in the extrapolation 
procedure. There are two important sources of instability: first, instabilities due to 
small variations of the central values  inside the error bars of what one may consider
the best fit. 
Secondly, we have the dependence of the results on the number of parameters used for the fits, 
or on the different functions used for the fits.
The way to deal with this is to try extrapolations with fits to various data sets (provided they
are compatible within errors), and to try fits with varying number of parameters 
and expressions. 
The final errors should be enlarged to encompass the results with these various fits: 
this is what will be done here.

\booksection{4. Results}
\vskip-0.5truecm
\booksubsection{4.1. Fits with only low energy (K decay) data}
 
\noindent
We first fit with only data from kaon decays, generically the more reliable ones. 
 In what follows, ``Old K decay data" refers to 
the data from K$_{l4}$ and K$_{2\pi}$ decays of 
 ref.~8  (for the exact 
value used from the  K$_{2\pi}$ analysis, see  Appendix~B 
here).
 ``NA48/2" denotes the 
K$_{l4}$ decay data of
ref.~9. For these last data statistical and systematic 
 errors are taken into account; they are added quadratically. 
 For the K$_{l4}$ data of Pislak et al.,
ref.~7,  we have added statistical and systematic errors in quadrature, with the exception of the
higher energy point, that we will discuss in Appendix~B.
Let us note that all  ``Old K decay data", as well as ``NA48/2" data,
   lie well inside the convergence region
of the  conformal expansion, as can be seen in \fig~2.  It
should also be
 noted  that  K$_{l4}$ decay data only give the difference
between the S0 and P wave phase shifts. However, this is not important as the 
P wave phase shift can be determined with great precision from the form factor of the
pion.\ref{10} Likewise,  K$_{2\pi}$ decays give the difference between S0 and S2 phases; but this
last is small on the kaon mass, and reasonably well known there.\ref{11}

\topinsert{
\setbox0=\vbox{\hsize 13truecm{\epsfxsize 11.truecm\epsfbox{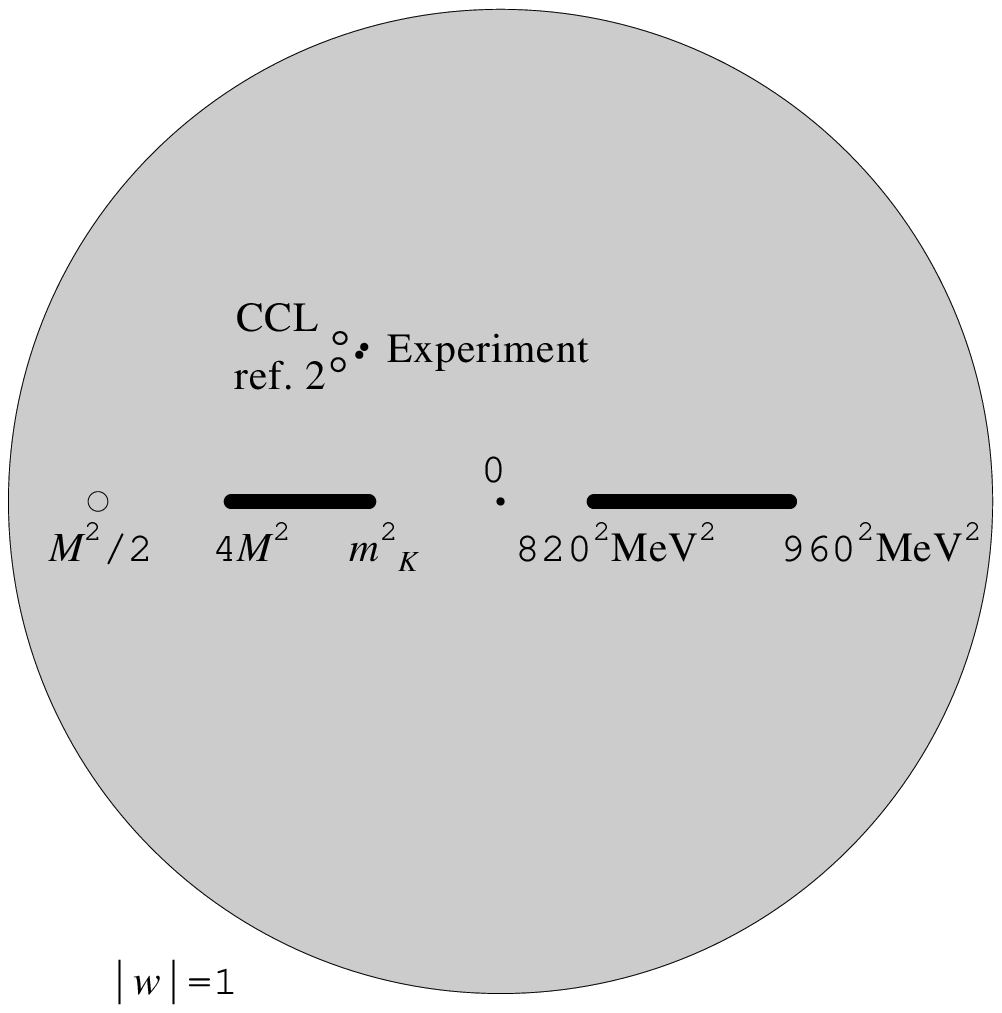}}}
\centerline{\box0}
\setbox6=\vbox{\hsize 12truecm\captiontype\figurasc{Figure 2. }{The $w$ disk, 
$|w|<1$. 
The thick lines are the regions 
where one has reliable experimental data. 
The sigma pole is also shown; the points labeled
``Experiment" are those obtained from our fits to experiment, 
Eqs.~(4.3) 
 and (4.10).
 The open point labeled CCL is the
Caprini, Colangelo, Leutwyler calculation,\ref{3} with error like the 
thickness of the point, and the open point labeled ref.~(2)
 is the average quoted in Eq.~(1.2).}}
\centerline{\box6}
}\endinsert
\bigskip

We start with only one parameter, $B_0$, plus the location of the zero, $\mu_0$,
in (3.1), or two parameters  $B_0$, $B_1$ in (3.2),
 if not separating the zero.\fnote{This second
method is less efficient  in the present case as 
the data region, $4M^2_\pi<s\leq m^2_K$, is very asymmetric with respect
to the origin 
$w=0$ around which we would be expanding.}  We first fit   Kaon decay data, 
including both  ``Old K decay data" and  ``NA48/2".
We find the following results ($a_0^{(0)}$ and $b_0^{(0)}$ 
are systematically given in units of $M_\pi$).

\smallskip\noindent
A.I) Separating the zero, i.e., with Eq.~(3.1): $\chidof=15.5/(22-2)$;
$$\eqalign{
B_0=&\,17.5\pm0.4,\quad\mu_0=744\pm24\;\mev\cr
a_0^{(0)}=&\,0.226\pm0.005,\quad b_0^{(0)}=0.287\pm0.008;
\cr
 M_\sigma=&\,493\pm5\;\mev,\quad \gammav_\sigma/2=228\pm9\;\mev.
}
\equn{(4.1)}$$

\smallskip\noindent
A.II) Not separating the zero, i.e., with Eq.~(3.2): $\chidof=15.8/(22-2)$;
$$\eqalign{
B_0=&\,5.1\pm0.3,\quad B_1=-18.7\pm0.6\cr
a_0^{(0)}=&\,0.221\pm0.006,\quad b_0^{(0)}=0.296\pm0.008;
\cr
 M_\sigma=&\,458\pm5\;\mev,\quad \gammav_\sigma/2=235\pm3\;\mev.
}
\equn{(4.2)}$$
The phase crosses $\pi/2$ at the energy $\mu_0=865\pm8\;\mev$.

It is clear that the errors in (4.1), (4.2) for the observable 
quantities, $a_0^{(0)}$, $b_0^{(0)}$, $M_\sigma$ and $\gammav_\sigma/2$,
 are purely statistical.
While  the results from (4.1) and (4.2) are similar, the differences 
are in general larger than these nominal errors.  
One can get a reasonable error estimate by averaging (4.1) and (4.2), 
and enlarging the errors by including a systematic uncertainty 
which amounts to half the difference between (4.1) and (4.2). In this way we get what 
we consider the best result using only K decay data,
$$\eqalign{
a_0^{(0)}=&\,0.224\pm0.006\;({\rm St.})\pm0.003\;({\rm Sys.}),
\quad b_0^{(0)}=0.292\pm0.008\;({\rm St.})\pm0.005\;({\rm Sys.});
\cr
 M_\sigma=&\,476\pm5\;({\rm St.})\pm18\;({\rm Sys.})\;\mev,\quad
\gammav_\sigma/2=234\pm9\;({\rm St.})\pm4\;({\rm Sys.})\;\mev. }
\equn{(4.3)}$$
These fits permit a good determination of the low energy parameters, 
but, as we will see later,
 are less reliable for extrapolation to the $\sigma$ pole;
nevertheless,  we still get  reasonable values for 
$M_\sigma$ and $\gammav_\sigma$.

An important question is the choice of parameters. We have elected to {\sl fix} 
the location of the Adler zero at $z_0=M_\pi$; but we could, for example in the fit 
separating the zero,  have fixed $\mu_0=810\,\mev$ 
(which is the approximate value we get from fits including data at higher energies, 
see below)
and consider, instead of it,  $z_0$ to be the free parameter. 
In this case we find
$$\eqalign{\mu_0\equiv&\,810\;\mev;\quad B_0=15.7\pm0.3,\quad z_0=174\pm10;
\cr
a_0^{(0)}=&\,0.222\pm0.009,\quad b_0^{(0)}=0.300\pm0.006;\cr
 M_\sigma=&\,476\pm5\;\mev,\quad
\gammav_\sigma/2=243\pm2\;\mev\cr
}
\equn{(4.4)}$$  
with $\chidof=15.6/(22-2)$: \equn{(4.4)}\ is quite compatible with 
(4.1), with an almost equal
\chidof\ For this reason, we will continue to fix $z_0=M_\pi$ in the remaining fits.

\booksubsection{4.2. Two parameter fits including also higher energy  data}

\noindent
 We denote by PY05-(2.13) to the set of combined data from various experiments, 
at energies $810\,\mev\leq s^{1/2}\leq952\,\mev$,
 collected in Eqs.~(2.13) in ref.~6 (and repeated here, 
Appendix~B, for ease of reference). Note that, as happened with the 
``K decay data",  these 
additional data also fall well inside the convergence region of the conformal
expansion (\fig~2).

\goodbreak

It is possible to 
fit all the K decay data (``Old" as well as ``NA48/2"),
 plus PY05-(2.13), with (3.1) and only one
$B_0$. We find the following results:

\smallskip
\noindent B.I) 
If separating the zero [with (3.1)],  $\chidof=21.5/(31-2)$ and 
$$\eqalign{
B_0=&\,17.2\pm0.4,\quad\mu_0=771\pm19\;\mev;\cr
a_0^{(0)}=&\,0.228\pm0.005,\quad b_0^{(0)}=0.286\pm0.007;
\cr
M_\sigma=&\,490\pm5\;\mev,\quad\gammav_\sigma/2=237\pm7\;\mev.\cr
}
\equn{(4.5)}$$

\smallskip
\noindent B.II)  {\sl Not} separating the zero at $\mu_0$, i.e, using (3.2), 
one finds  $\chidof=21.3/(31-2)$ and
$$\eqalign{
B_0=&\,2.8\pm0.3,\quad B_1=-24.0\pm0.6;\cr
a_0^{(0)}=&\,0.213\pm0.006,\quad b_0^{(0)}=0.296\pm0.008;
\cr
M_\sigma=&\,467\pm5\;\mev,\quad\gammav_\sigma/2=213\pm3\;\mev.\cr
}
\equn{(4.6)}$$

These fits with only two parameters are somewhat rigid; fortunately,
 it is possible to include in
the fit one parameter more, 
that permits  more  flexibility.

\booksubsection{4.3. Three parameter fits including also higher energy  data}

\noindent

 When we include the  PY05-(2.13)
together with all (old as well as NA48/2)  K decay data, it is possible to
include a further  parameter in the fit, thus making it more flexible,  getting also a better
\chidof; this was not possible with 
K decay data alone, since there would have been too few 
data and including a superfluous parameter would 
have given spurious minima. Moreover, the fits will now be   more
realistic  because the terms in 
$B_1$, $B_2$ represent (in the average) the cuts of the effective range function,
 and because now the
data are more symmetrically distributed, lying on both sides of 
the point around which we are expanding, $w=0$; see \fig~2.

We fit as in the previous subsection and find the following results.
\smallskip
\noindent C.I)   If we use (3.1), i.e., we separate off the zero, 
the \chidof\ becomes $\chidof=21.0/(31-3)$ and we get 
$$\eqalign{B_0=&\,19.0\pm0.4,\quad B_1=4.4\pm0.8,\quad
 \mu_0=781\pm19\;\mev;\cr
a_0^{(0)}=&\,0.235\pm0.008,\quad b_0^{(0)}=0.282\pm0.008;\cr 
M_\sigma=&\,496\pm6\;\mev,\quad \gammav_\sigma/2=258\pm8\;\mev.\cr
..
\cr}
\equn{(4.7)}$$
\smallskip
\noindent 

C.II) Not separating the zero, i.e., with (3.2): the \chidof\ improves 
significantly 
when introducing the new 
parameter, $B_2$. We find $\chidof=18.7/(31-3)$ and 

$$\eqalign{
{B}_0=&\,4.3\pm0.3,\quad {B}_1=-26.7\pm0.6,\quad
{B}_2=-14.1\pm1.4;
\cr 
 a_0^{(0)}=&\,0.231\pm0.009,\quad b_0^{(0)}=0.288\pm0.009;\cr
M_\sigma=&\,474\pm6\;\mev\quad
 \gammav_\sigma/2=254\pm4\;\mev.\cr
}
\equn{(4.8)}$$
The zero of $\cot\delta_0^{(0)}(s)$ now occurs at 
$s=\mu^2_0$, with $\mu_0=801\pm6\,\mev$.

\topinsert{
\bigskip
\setbox0=\vbox{{\psfig{figure=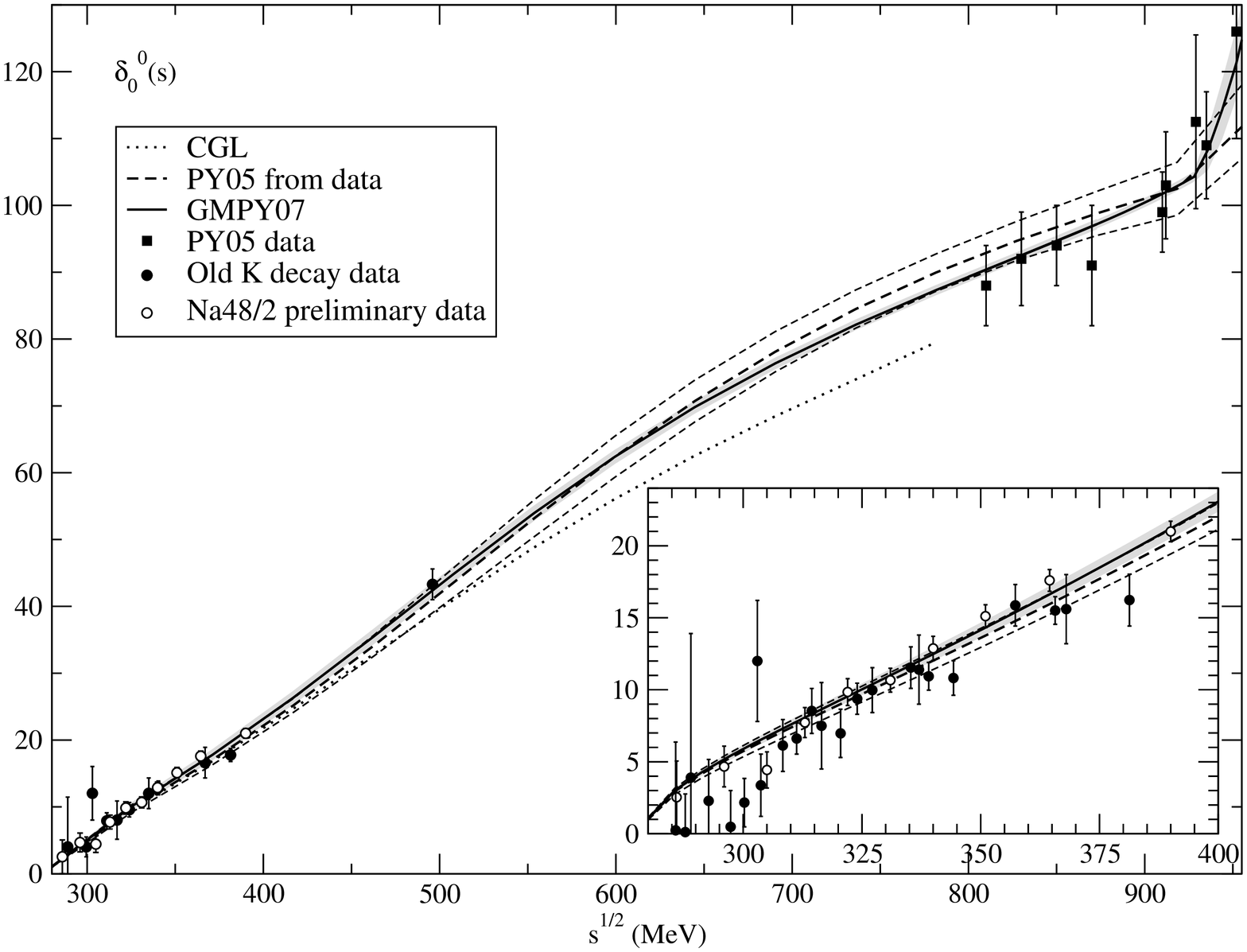,width=14.0truecm,angle=-90}}}
\centerline{\box0}
\setbox6=\vbox{\hsize 12truecm\captiontype\figurasc{Figure 3. }{The data 
points fitted, together with the phase shifts corresponding to 
our best fit, (4.8) [full line, with error given by the dark band]. 
Dashed lines: the error band of the ``old" PY05 fit in ref.~6. The dotted line is the prediction of
ref.~7, based on  Roy equations and the value (1.6) for $a^{(0)}_0$.
In the inserted box, a blow-up of the low energy region.}}
\centerline{\box6}
}\endinsert
\bigskip

In principle, (4.8), shown in \fig~3,
 is the best fit: it takes into account the most of 
theoretical information, and it has the smallest \chidof\ 
However, (4.7) also takes into account the same 
theoretical information, and its \chidof, while larger, is quite comparable 
to that of (4.8); 
in fact, the phase shifts described by (4.7) and (4.8) 
overlap within their errors.

Before continuing, a few words have to be said on the errors in (4.8) [or (4.7)].
The level of precision of these fits is such that the errors
 are comparable to the expected contributions of 
isospin symmetry breaking,  
 which should therefore be considered. 
For the best values of the low energy parameters and 
location of the $\sigma$~pole, this is not important; we will 
compose the results in (4.7) and (4.8), adding as systematic error 
half the difference, and the final uncertainty 
is substantially larger than the estimated effect of isospin breaking corrections; 
for the {\sl parametrization}, we will discuss this  in \sect~6.  

As we did for the fits with only K decay data, we average the results 
in (4.7) and (4.8), weighted now each with its corresponding 
\chidof, and enlarge the errors including a systematic error of half the difference. 
In this way we find what we consider our final result, for the low energy parameters and 
location of the $\sigma$~pole:

$$\eqalign{
a_0^{(0)}=&\,0.233\pm0.010\;({\rm St.})\pm0.003\;({\rm Sys.})\;M^{-1}_\pi,\quad
b_0^{(0)}=0.285\pm0.009\;({\rm St.})\pm0.003\;({\rm Sys.})\;M^{-3}_\pi;\cr
M_\sigma=&\,484\pm6\;({\rm St.})\pm11\;({\rm Sys.})\;\mev,\quad
\gammav_\sigma/2= 
255\pm8\;({\rm St.})\pm2\;({\rm Sys.})\;\mev.
\cr
}
\eqno{(4.9)}$$

Eqs.~(4.9) and  (4.3) are  {\sl very} compatible, showing the stability of the fits, 
and of the extrapolations to the sigma pole, against the number of parameters used 
and the formulas employed. The best values for the complex zero, $\bar{s}_\sigma$ are shown,
in the $w$-plane, in
\fig~2, where  we have collected the results of our fits
with two
parameters to K decay data only and their average, together with
our three parameters fits.
The results of our best estimate and the more representative of our results for the $\sigma$ pole 
and low energy parameters 
 are collected  in the accompanying 
Table ($a_0^{(0)}$ and $b_0^{(0)}$ 
in units of $M_\pi$).
\medskip
\setbox9=\vbox{\petit
$$\matrix{\hbox{Process}\vphantom{\Bigg|} &M_\sigma\;[\mev]&\gammav_\sigma/2\;[\mev]
&\phantom{XX}&a_0^{(0)}&b_0^{(0)}\cr
\hbox{K decay data, 2 par., (4.1)} & 493\pm5 & 228\pm9 & & 0.226\pm0.005 & 0.287\pm0.008 \cr
\hbox{K decay data, 2 par., (4.2)} & 458\pm5 & 253\pm3 & & 0.221\pm0.006 & 0.296\pm0.008 \cr
\vphantom{\Bigg|}
\hbox{K decay data average} & 476\pm23 & 234\pm13 & & 0.224\pm0.009 & 0.292\pm0.013 \cr
\hbox{K decay data + PY05, 3 par., (4.7)} & 496\pm6 & 258\pm8 & & 0.235\pm0.008 & 0.282\pm0.008 \cr
\hbox{K decay data + PY05, 3 par., (4.8)}& 474\pm6 & 254\pm4 & & 0.230\pm0.010 & 0.288\pm0.009 \cr
\vphantom{\Bigg|}
\hbox{\bf Best estimate} & {\bf 484\pm17} & {\bf 255\pm10} & & {\bf 0.233\pm0.013} & {\bf 0.285\pm0.012} \cr
}
$$}
\centerline{\tightboxit{\box9}}
\smallskip

One may wonder whether our choice of data at high energies may have biased our results.
To test this,  we will consider two alternate fits, using, instead of the PY-05 
collection of data, the 
 K decay data plus the data sets B, C given by Grayer et al.\ref{12}
We choose these solutions out of the four solutions in ref.~12 because 
they are the ones that 
satisfy best forward dispersion relations [apart, of course, 
from our solution (4.8) here], as shown in ref.~6.
 We find $\chidof=55.7/(48-3)$ and
$$\eqalign{
&\hbox{[K decay data plus
Sol. C of Grayer et al.\ref{12}]}:\cr
B_0=&\,3.57\pm0.17,\quad B_1=-24.3\pm0.5,\quad B_2=-6.3\pm1.3;\cr
a_0^{(0)}=&\,0.226\pm0.008,\quad b_0^{(0)}=0.299\pm0.007;\cr
M_\sigma=&\,465\pm5\;\mev,\quad \gammav_\sigma=231\pm3\;\mev.\cr
}
\equn{(4.10a)}$$
For solution~B, we have $\chidof=31.4/(41-3)$ and
$$\eqalign{
&\hbox{[K decay data plus
Sol. B of Grayer et al.\ref{12}]}:\cr
B_0=&\,7.63\pm0.23,\quad B_1=-23.2\pm0.6,\quad B_2=-23.0\pm1.4;\cr
a_0^{(0)}=&\,0.251\pm0.011,\quad b_0^{(0)}=0.269\pm0.006;\cr
M_\sigma=&\,477\pm7\;\mev,\quad \gammav_\sigma=321\pm6\;\mev.\cr
}
\equn{(4.10b)}$$
It should be noted that the errors in (4.10a,~b) 
are purely nominal; in fact, 
the errors in the Solutions~B,~C of Grayer et al. that we have used are only the 
{\sl statistical} errors: 
while the very existence of several solutions to the 
same set of raw data in ref.~12 shows that {\sl systematic} 
errors must be substantially larger (by a factor 3 or more; see the 
comments at the end of Appendix~B, and Fig.~4 here). 
This is reflected in the incompatibility of the fits. 
However, and
in spite of these problems with the errors, 
these new fits show the stability of the results for $a_0^{(0)}$, $b_0^{(0)}$
and, to a lesser extent, $M_\sigma$; while the value of $\gammav_\sigma$ is 
shown to be less reliably determined.

\topinsert{
\bigskip
\setbox0=\vbox{{\psfig{figure=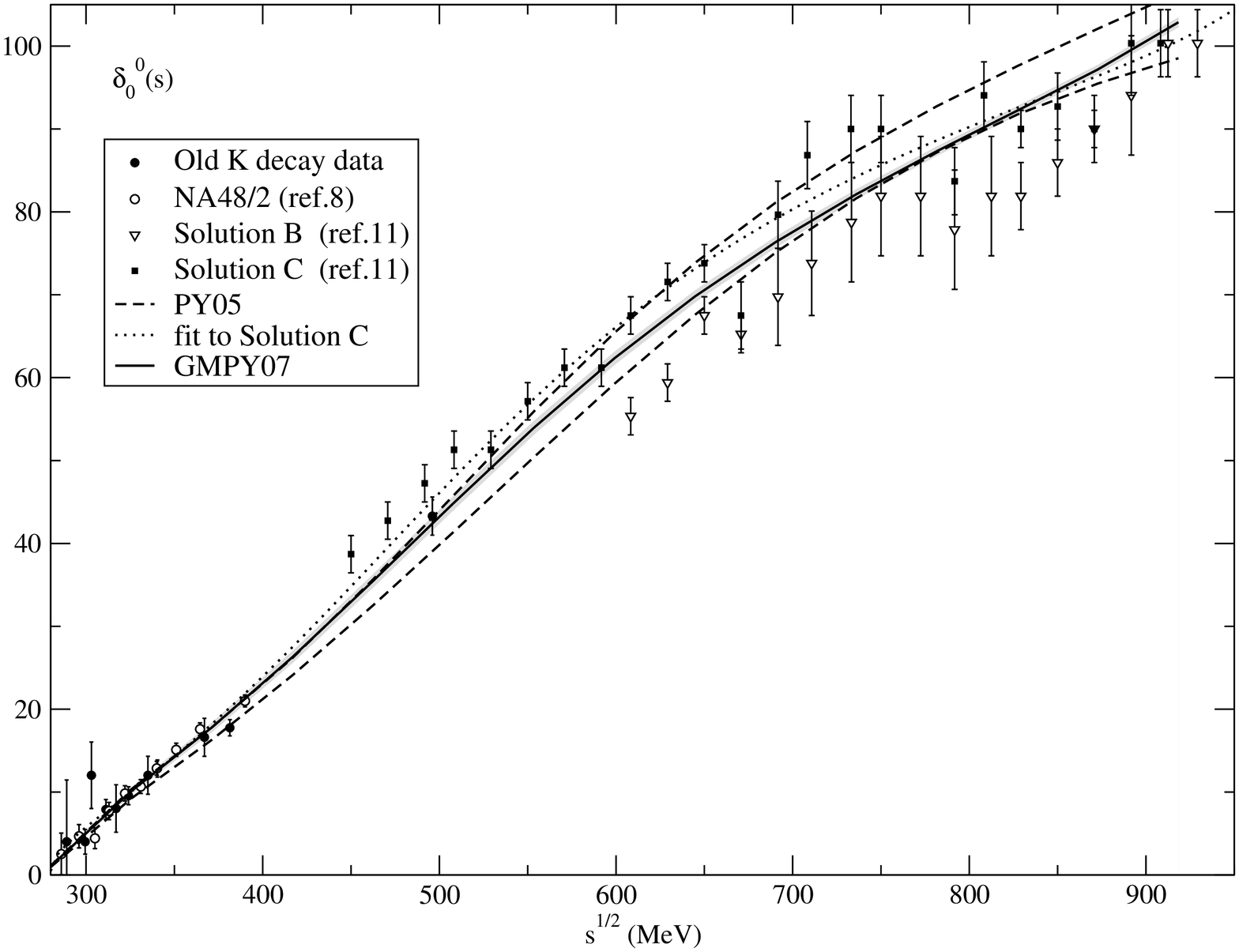,width=14.0truecm,angle=-90}}}
\centerline{\box0}
\setbox6=\vbox{\hsize 12truecm\captiontype\figurasc{Figure 4. }{The fit to the 
data in Solution~C of Grayer et al. [Eq.~(4.10), dotted line], together with the error band for the
``old" fit in PY05 (dashed lines), and our best fit here [Eq.~(4.8), continuous line and gray band].
The
data for  Solutions B and C of Grayer et al. are also shown. }}
\centerline{\box6}
}\endinsert
\bigskip

Finally, one may also wonder about the possibility of 
adding more parameters to the fits. 
It turns out that a fourth parameter would be redundant, as the 
\chidof\ does not improve appreciably by introducing it and spurious minima appear.
Specifically, we find $\chidof=17.5/(31-4)$ and 
the central values for the parameters ${B}_0=5.8$, ${B}_1=-22.4$, 
${B}_2=-26.3$  
and ${B}_3=-32.5$, which do not decrease at the expected rate.

\booksection{5. The scattering length with the effective range expansion}

\noindent
The scattering length (but of course, not the location of the $\sigma$ pole) 
can also be obtained with the effective range expansion. We 
expand  the effective range function in powers of $k^2$,
$$\psi(s)={{1}\over{M_\pi a_0^{(0)}}}+4R_0k^2/M^4_\pi+\cdots;
\equn{(5.1)}$$
only two terms are necessary. The expansion here, however, is poorly   convergent 
when compared to the conformal mapping one;
due to the presence of the
Adler zero at $s=M_\pi^2/2$, the circle of convergence only extends to 
$s=\tfrac{15}{2}M^2_\pi$, i.e., to energies $s^{1/2}\lsim385\,\mev$.
 To remain well inside the region of convergence of 
(5.1), we only fit 
data at energies $s^{1/2}\leq351\,\mev$, and  one then finds
($a_0^{(0)}$ and $R_0$ in units of
$M_\pi$)
$$\eqalign{
\chi^2/{\rm d.o.f.}=\dfrac{6.3}{9-2};&\,
\quad a_0^{(0)}=0.225\pm0.007^{+0}_{-0.15},\quad R_0=-[1.00\pm0.08^{+0.27}_{-0}];
\quad\hbox{[Old K
decay data\ref{8}]}.\cr
\chi^2/{\rm d.o.f.}=\dfrac{2.4}{7-2};&\,
\quad a_0^{(0)}=0.229\pm0.005\pm0.024,\quad
R_0=-[0.93\pm0.05^{+0.38}_{-0.22}];\quad\hbox{[NA48/2 data\ref{9}]}\cr
\chi^2/{\rm d.o.f.}=\dfrac{8.8}{17-2};&\,
\quad a_0^{(0)}=0.229\pm0.004\pm0.021,\quad R_0=-[0.94\pm0.04^{+0.34}_{-0.21}];
\quad\hbox{[Old K decay+NA48/2]}.\cr
}
\equn{(5.2)}$$
The first  errors here are statistical; 
the second are obtained   changing the point where one stops fitting 
to 367~\mev\ or to 340~\mev. 
Although the results are compatible with 
(3.7), it is clear that the effective range method is much less precise and also less 
stable than the conformal mapping one --not surprisingly, as the last incorporates a lot more of
analyticity information.

\booksection{6. Summary and discussion}

\noindent
In summary: in the present paper we have found an accurate representation of the 
S0 wave, given by the parametrization (4.8), that we consider the best of our results. 
However, and as noted just after this equation, the level of precision 
attained requires that we consider isospin breaking effects 
related, for example, to whether we use the parametrization for 
$\pi^+\pi^-$ or $\pi^0\pi^0$ 
scattering. 

It turns out that electromagnetic corrections, which are likely the largest part 
of isospin breaking corrections, are taken into account in the analysis of {\sl systematic} 
errors in the NA48/2 experiment. This is the only place where they can be important, as the errors in
all other 
experiments are much larger than the estimated isospin breaking effects; for example,
  for the P wave, where separate fits were made in 
ref.~10 to $\pi^0\pi^+$ and $\pi^+\pi^-$ processes, 
the corresponding corrections were found to be at the 1\% level, except for the 
kinematical corrections. We therefore take this into account 
by considering that, in the  parametrization in (4.8), that we repeat here for ease of reference, 
$$\eqalign{
\cot\delta_0^{(0)}(s)=&{{s^{1/2}}\over{2k}}\,{{M^2_\pi}\over{s-\tfrac{1}{2}z_0^2}}\,
\Big\{\dfrac{z_0^2}{M_\pi\sqrt{s}}+{B}_0+{B}_1w(s)+{B}_2w(s)^2\Big\},\quad
w(s)=\dfrac{\sqrt{s}-\sqrt{4m^2_K-s}}{\sqrt{s}+\sqrt{4m^2_K-s}};\cr
{B}_0=&\,4.3\pm0.3,\quad {B}_1=-26.7\pm0.6,\quad
{B}_2=-14.1\pm1.4;\quad z_0=M_\pi,\cr
}
\equn{(6.1a)}$$
 we have to interpret
$$k=\cases{
\sqrt{s/4-M_\pi^2},\quad M_\pi=m_{\pi^+}=139.57\;\mev\;{\rm for}\; \pi^+\pi^-\;{\rm scattering},\cr
\sqrt{s/4-M_\pi^2},\quad M_\pi=m_{\pi^0}=134.98\;\mev\;{\rm for}\; \pi^0\pi^0\;{\rm scattering},\cr
}\equn{(6.1b)}$$
according to the process for which it is used.

The  values for scattering length, effective
 range parameter and location of the sigma pole from (6.1)  are
$$\eqalign{
a_0^{(0)}=&\,0.231\pm0.009,\quad b_0^{(0)}=0.288\pm0.009;\cr
M_\sigma=&\,474\pm6\;\mev\quad
 \gammav_\sigma/2=254\pm4\;\mev.\cr
}
\equn{(6.2)}$$
This is perfectly compatible, both in central values and errors, with the 
results reported in (4.9). Because the numbers in (4.9) include a systematic error due to 
comparison of results with two fitting formulas (something that is of course impossible
to do for the 
parametrization itself), they
 should be considered to be the more reliable results, for these quantities; so
we repeat them here for ease of reference:
$$\eqalign{
a_0^{(0)}=&\,0.233\pm0.010\;({\rm St.})\pm0.003\;({\rm Sys.})\;M^{-1}_\pi,\quad
b_0^{(0)}=0.285\pm0.009\;({\rm St.})\pm0.003\;({\rm Sys.})\;M^{-3}_\pi;\cr
M_\sigma=&\,484\pm6\;({\rm St.})\pm11\;({\rm Sys.})\;\mev,\quad
\gammav_\sigma/2= 
255\pm8\;({\rm St.})\pm2\;({\rm Sys.})\;\mev.
\cr
}
\eqno{(6.3)}$$
These results overlap, within errors, with what was found in ref.~7 
using chiral perturbation theory and analyticity. 
The value of $a_0^{(0)}$ is also compatible with experimental determinations\ref{13} 
of this quantity, using a method devised by Cabibbo from $K\to\pi^+\pi^0\pi^0$ 
decays, or from pionium decay\ref{14} 
[although our results are more precise and do {\sl not} depend on knowledge of 
$a_0^{(2)}$].

Another question is whether one can improve our determinations. 
The answer is no, if we only use experimental data on the S0 wave. 
But it is possible to give an independent, perhaps more precise determination,
 calculating 
the function $\cot\delta_0^{(0)}(s)$ for complex $s$ with dispersion relations 
(Roy equations).
 This
necessitates input of experimental data on other waves,  and input in the
 high energy, Regge
region, so it is not a determination based purely on S0 wave experimental data. 
Preliminary results indicate that one would find numbers for 
$M_\sigma$  and $\gammav_\sigma/2$ compatible with (1.3), and with smaller errors.
This procedure could also  give  improved values for the scattering length 
and effective range parameters.

\booksection{Appendix A: On ghosts}

\noindent
As remarked in the main text, we have separated off a term $z^2_0/M_\pi\sqrt{s}$ 
in Eqs.~(3.1), (3.2) to avoid the appearance of a ghost, i..e., of a spurious 
pole in $\hat{f}_0^{(0)}(s)$ located between the Adler zero and $s=0$. 
It turns out that the effect of this ghost is 
 is negligible. To show this, we  
 repeat our best fit, Eq.~(4.8), but not eliminating the ghost; 
that is to say, we replace Eq.~(3.2) 
by 
$$\eqalign{
\cot\delta(s)=&{{s^{1/2}}\over{2k}}\,{{M^2_\pi}\over{s-\tfrac{1}{2}z_0^2}}\,
\big\{\hat{B}_0+\hat{B}_1w(s)+\hat{B}_2w(s)^2+\cdots\big\},\cr
}
\equn{(A.1)}$$
i.e., without the term that removes the ghost.  
Then we
find  (4.8) replaced by
$$\eqalign{
\hbox{With Eq. (A.1):}\quad \dfrac{\chi^2}{{\rm
dof}}=\dfrac{18.8}{31-3};\quad \hat{B}_0&\,=4.5\pm0.3,
\quad \hat{B}_1=-26.9\pm0.6,\quad
\hat{B}_2=-13.5\pm1.4;\cr 
 a_0^{(0)}=&\,0.231\pm0.010;\quad  b_0^{(0)}=0.287\pm0.008;
\cr
  M_\sigma=&\,475\pm6\;\mev\quad
 \gammav_\sigma/2=253\pm5\;\mev;\cr
}
\equn{(A.2)}$$
this corresponds to $\mu_0=801\pm6\,\mev$, 
and is practically indistinguishable from (4.8). 
Removing the ghost is  little more than an aesthetical 
 requirement.

\booksection{Appendix B: On experimental data}

\noindent
We here say a few words have  about the 
error we have taken for the datum at highest energy of Pislak et al.\ref{8}
In this reference, the number given is
$$\delta_0^{(0)}((381.4\;\mev)^2)-\delta_1((381.4\;\mev)^2)=
0.285\pm0.014\,({\rm St.})\pm0.03 \,({\rm Sys.})$$
Now, this result is suspicious. The error is the smallest of
 all those among the data of Pislak
et al.,\ref{8} although the datum is actually an average 
value near the edge of phase space.
 Moreover, as we will see in a moment, the central value 
is incompatible with other
determinations.  For this reason, we have, in our fits, done as in ref.~6
 and multiplied the statistical error by a
factor 1.5: so, we have taken
$$\delta_0^{(0)}((381.4\;\mev)^2)-\delta_1((381.4\;\mev)^2)=
0.285\pm(1.5\times0.014).$$
We could have chosen, instead of this, to add systematic and statistical error 
{\sl linearly} for this datum, i.e., to fit with
$$\delta_0^{(0)}((381.4\;\mev)^2)-\delta_1((381.4\;\mev)^2)=
0.285\pm0.017.$$
In this case, (4.8) is replaced by
$$\eqalign{
\dfrac{\chi^2}{{\rm
dof}}=\dfrac{20.0}{31-3};&\,\quad {B}_0=4.53\pm0.27,\quad {B}_1=-26.9\pm0.6,\quad
{B}_2=-15.3\pm1.4;
\cr 
 a_0^{(0)}=&\,0.231\pm0.010,\quad b_0^{(0)}=0.285\pm0.008;\quad \mu_0=804\pm5\;\mev;\cr
M_\sigma=&\,476\pm6\;\mev\quad
 \gammav_\sigma/2=258\pm5\;\mev,\cr
}
$$
that is to say, almost identical to (4.8). 

To ascertain the degree of incompatibility of the datum of 
Pislak et al. with the others, we have repeated the fits with two other possibilities: 
first, adding its errors in quadrature. In this case, the $\chi^2$ increases to 21.6, 
i.e., three units above what we got in (4.8). Alternatively,
if we {\sl remove} the datum from
the fit, 
 the $\chi^2$ decreases to 15.7: the datum  at  381.4~\mev\
of  Pislak et al., with its original error, carries a penalty of increase
of the $\chi^2$  by  almost six units and would certainly 
 bias the results if included {\sl tel quel}.
 This   justifies out treating its error as we have
done.

Next, we explain the value we give for the datum at $m_K^2$. 
From the decays $K^+\to \pi^0\pi^+$,  $K_S\to \pi^0\pi^0$ 
and  $K_S\to \pi^+\pi^-$ one can obtain the 
difference $\delta_0^{(0)}(m^2_K)- \delta_0^{(2)}(m^2_K)$. 
{\sl Neglecting} radiative corrections,
and with the latest results from Kloe\fnote{We are grateful to Dr. C.~Gatti for
comunicating us these results before formal publication.}
one finds
$$\delta_0^{(0)}(m^2_K)- \delta_0^{(2)}(m^2_K)=51.27\pm0.82\degrees.$$
{\sl Including} radiative corrections as in  ref.~15, this is corrected to
$$\delta_0^{(0)}(m^2_K)- \delta_0^{(2)}(m^2_K)=57.27\pm0.82\;
 ({\rm exp}) \pm3\; ({\rm radiative}) 
\pm1\degrees\; ({\rm chiral\; perturbation\; approximations}).$$
Subtracting the value of $\delta_0^{(2)}(m^2_K)$ obtained from fits to 
experiment in ref.~6, and adding the errors linearly 
(as is advisable given the uncertainties in their evaluation) we arrive 
at the result we have been using in our fits:
$$\delta_0^{(0)}(m^2_K)=48.7\pm4.9\degrees.
\equn{(B.1)}$$
We have verified that a very similar result is found if using the PDT\ref{1} 
decay data, and the prescription for radiative corrections of refs.~16.

Finally, and for ease of reference, we repeat here the data PY05 [ref.~6, Eqs. (2.13)] used in
the fits:
$$\eqalign{
\delta_0^{(0)}(0.870^2\,\gev^2)=&\,91\pm9\degrees;\quad
\delta_0^{(0)}(0.910^2\,\gev^2)=\,99\pm6\degrees;\quad
\delta_0^{(0)}(0.935^2\,\gev^2)=109\pm8\degrees;\cr
\delta_0^{(0)}(0.912^2\,\gev^2)=&\,103\pm8\degrees;\quad
\delta_0^{(0)}(0.929^2\,\gev^2)=112.5\pm13\degrees;\quad
\delta_0^{(0)}(0.952^2\,\gev^2)=126\pm16\degrees;\cr
\delta_0^{(0)}(0.810^2\,\gev^2)=&\,88\pm6\degrees;\quad
\delta_0^{(0)}(0.830^2\,\gev^2)=92\pm7\degrees;\quad
\delta_0^{(0)}(0.850^2\,\gev^2)=94\pm6\degrees.
}
\equn{(B.2)}$$ 
Note that the errors here include systematic errors, 
estimated by comparing different determinations;
 they are a factor 3 or more larger than the
nominal, statistical errors of the different phase shift analyses (e.g., those in 
ref.~12).

\vfill\eject
\booksection{Acknowledgments}

\noindent  Part of the results in this note have been
presented by 
 J. R. Pel\'aez at the NA48/2 meeting in Geneva, December 2006. 
We are  grateful to Dr. Brigitte~Bloch-Devaux for
communicating us the  phase shifts obtained in the NA48/2 analysis,\ref{9} 
including the (as yet unpublished) analysis of the 
systematic errors; 
as well as to C. Gatti for communicating us the $K\to\pi\pi$ results
before formal publication.
One of us (FJY) also would like to thank  Dr. Caprini 
for very useful correspondence. 
Finally, discussions with P.~Minkowski and W.~Ochs have been very helpful 
to make us understand the
interest of discussing the solutions~B,~C of Grayer et al.

This work was supported in part by the Spanish DGI of the MEC under
contract FPA2003-04597.
RGM and JRP research was partially funded by Spanish CICYT contracts
FPA2005-02327, as well as Banco Santander/Complutense
contract PR27/05-13955-BSCH. JRP research is
also partially funded by  CICYT contract FIS2006-03438 
and is part of the EU integrated
infrastructure initiative HADRONPHYSICS PROJECT,
under contract RII3-CT-2004-506078.

\booksection{References}

\item{1 }{PDT: Eidelman,  S., et al., {\sl Phys. Letters} {\bf B592}, 1 
 (2004).}
\item{2 }{Pel\'aez, J. R., {\sl Modern Phys. Lett.~A}, {\bf 19}, 2879 (2004);
Oller, J. A. , Oset, E., and  Pel\'aez, J.
R., {\sl Phys. Rev.} {\bf D59}, 074001 (1999) and 
Erratum, {\sl ibid.} {\bf D60}, 099906 (1999); A. Dobado and Pel\'aez, J. R., 
 {\sl Phys. Rev.} {\bf D56}, 3057 (1997); G\'omez-Nicola,~A., and
 Pel\'aez, J. R., {\sl Phys. Rev.} {\bf D65}, 054009 (2002).}
\item{3 }{Caprini, I., Colangelo, G., and Leutwyler, H. {\sl  Phys. Rev. Lett.} 
{\bf 96}, 132001 (2006).}
\item{4 }{Zhou, Z. Y., 
et al. {\sl JHEP} 0502, 043 (2005).}
\item{5 }{Girlanda,~L., et al.,
 {\sl Phys. Lett.}   {\bf B409}, 461 (1997);
Descotes-Genon,~S., et al.,
{\sl  Eur. Phys. J.} {\bf C24}, 469 (2002);
Pelaez,~J.~R. and Yndurain,~F.~J.,
{\sl  Phys. Rev.} {\bf D68}, 074005 (2003);
Pelaez,~J.~R. and Yndurain,~F.~J.,
{\sl  Phys. Rev.} {\bf D69}, 114001 (2004);
Yndurain,~F.~J.,
{\sl  Phys. Lett.}   {\bf B612}, 245 (2005) and
  arXiv: hep-ph/0510317.}
\item{6 }{Pel\'aez, J. R., and Yndur\'ain,~F.~J., {\sl Phys. Rev.} {\bf D71}, 074016
(2005).}
\item{7 }{Colangelo, G., Gasser, J.,  and Leutwyler, H.,
 {\sl Nucl. Phys.} {\bf B603},  125 (2001).}
\item{8 }{{\sl Kl4 decays}\/: Rosselet, L., et al. {\sl Phys. Rev.} {\bf D15}, 574  (1977); 
Pislak, S.,  et al.  {\sl
Phys. Rev. Lett.}, {\bf 87}, 221801 (2001).
{\sl $K\to2\pi$ decays}\/: Aloisio, A., et al., {\sl Phys. Letters}, {\bf B538}, 21  (2002).}
\item{9 }{NA48/2 ( CERN/SPS experiment); 
Bloch-Devaux, B., presented at QCD06 in Montpellier
(France), 3-7 July 2006 and
Masetti, L.,  presented at ICHEP06 in Moscow (Russia), 26 July to 2 August 2006.}
\item{10}{de Troc\'oniz, J. F., and Yndur\'ain, F. J., {\sl Phys. Rev.},  {\bf D65},
093001,
 (2002),  
and {\sl Phys. Rev.} {\bf D71}, 073008 (2005).} 
 \item{11}{Losty, M.~J., et al.  {\sl Nucl. Phys.}, {\bf B69}, 185 (1974); 
Hoogland, W., et al. 
{\sl Nucl. Phys.}, {\bf B126}, 109 (1977); 
Durusoy,~N.~B., et al., {\sl Phys. Lett.} {\bf B45}, 517 (1973). 
A global fit may be found in ref.~5.}
\item{12}{Grayer, G., et al., (1974). {\sl Nucl. Phys.}  {\bf B75}, 189.}
\item{13}{Cabibbo, N., {\sl Phys.Rev.Lett.} {\bf 93}: 121801 (2004); 
Cabibbo, N., and Isidori, G., {\sl JHEP} 0503:021 (2005); Batley,~J.~R., et al.,
{\sl Phys. Letters} {\bf B633}, 173 (2006).}
\item{14}{Adeva, B., et al. {\sl Phys.Lett.} {\bf B619}, 50 (2005).}
\item{15}{Cirigliano, V., et al, {\sl Eur. Phys. J.} {\bf C33}, 369 (2004).} 
\item{16}{Nachtmann, O, and de Rafael, E. (1969). CERN preprint TH-1031 (unpublished); 
Pascual, P., and Yndur\'ain, F. J.  {\sl Nucl. Phys.} {\bf B83}, 362 (1974).} 

\bye